\documentclass[aps]{revtex4}%
\usepackage{amsfonts}
\usepackage{graphicx}
\usepackage{pifont}
\usepackage{graphicx}
\usepackage{amsmath,amssymb,amsthm}
\usepackage{amsmath}
\usepackage{amssymb}%
\providecommand{\U}[1]{\protect\rule{.1in}{.1in}}
\newcommand{\ini}{\begin{equation}}
\newcommand{\fin}{\end{equation}}
\newcommand{\inia}{\begin{eqnarray}}
\newcommand{\fina}{\end{eqnarray}}
\begin{document}
\title{\textbf{OPERA superluminal neutrinos explained by spontaneous emission and
stimulated absorption.}}
\author{Rafael S. Torrealba\ S. \thanks{rtorre@ucla.edu.ve; rtorre@uicm.ucla.edu.ve}}
\affiliation{Departamento de F\'{\i}sica.Universidad Centro Occidental "Lisandro Alvarado"}

\begin{abstract}
In this work it is shown, that for short $3\mathit{ns}$ neutrino pulses
reported by OPERA, a relativistic shape deforming effect of the neutrino
distribution function due to spontaneous emission, produces an earlier arrival
of $65.8\mathit{ns}$ in agreement with the reported $62.1\mathit{ns\pm
}~3.7\mathit{ns}$, with a RMS of $16.4\mathit{ns}$ explaining the apparent
superluminal effect. It is also shown, that early arrival of long
$10500\mathit{ns}$ neutrinos pulse to Gran Sasso, by $57.8\mathit{ns}$ with
respect to the speed of light, could be explained by a shape deforming effect
due to a combination of stimulated absorption and spontaneous emission, while
traveling by the decay tunnel that acts as a LASER tube.


\end{abstract}
\maketitle



\section{Introduction}

Recently, OPERA collaboration has reported 20 superluminal neutrinos coming
from $3\mathit{ns}$ LHC extractions, in addition to the $16000$ neutrinos,
collected for 3 years coming from $10500\mathit{ns}$ LHC\ extractions
\cite{OPERA} that apparently are faster than light for $\approx60\mathit{ns}$
while traveling 730 km from CERN to Gran Sasso National Laboratory. What is
most amazing is that the velocity obtained is even faster than the known speed
of light in the vacuum, within margins of error accurately calculated. This
result is in direct contradiction with Einstein's Relativity Theory that has
been the angular corner of modern physics for more than a century. The speed
of light in vacuum is considered to be the limit velocity for all matter and
information travel, as a consequence of the causality principle.

OPERA faster than light neutrinos is in direct contradiction with astronomical
observations as the SuperNova 1987A \cite{1987A}: in 1987 three independent
neutrino observatories: \textit{Kamiokande II} in Japan, \textit{IMB} in USA,
and \textit{Baksan} in the former USSR detected 11, 8 and 5 neutrino events,
in a burst lasting less than $13\mathit{s}$ almost at the same UTC time.
Approximately three hours later the light from the Nova was observed. This
does not indicate that neutrinos arrived faster than light, the accepted
explanation is that neutrinos arise from the collapse of the star core, but
the burst of light occurs only when the shock waves reach the star surface. If
OPERA result is right, the neutrinos must precede the light by $60\mathit{ns}$
for each $730\mathit{km}$, as the Nova 1987A is 168.000 light year away from
the earth, neutrinos must arrived earth 1500 days BEFORE the explosion was
observed, an not few hours earlier.

Even recently, it has been argued that superluminal neutrino will decay very
fast by Cherenkov analogue effect due to neutral current
interactions\cite{Glashow}. ICARUS collaboration, another team at Gran Sasso,
had reported that superluminal CNGS neutrinos do not decay as theoretically
expected \cite{ICARUS} questioning OPERA result. To my point of view this
objection to OPERA results is not valid because is based in a relativistic
model, that will be not longer valid if superluminal neutrinos do exist. It is
the same as to said that Bohr atom is forbidden because of Larmor formula will
make the electrons to radiate.

In a recent work \cite{Stimulated}, it has been proposed an explanation for
early arrival by $57.8\mathit{ns}$ of the neutrinos with respect to the speed
of light coming from "long" $10500\mathit{ns}$ LHC\ extractions. This could be
explained by a shape deforming effect of the neutrino distribution function,
with respect to the proton distribution function (PDF) due to the stimulated
absorption. But in this work it was also probed that stimulated absorption can
not explain the apparent earlier arrival of 20 neutrinos reported for "short"
$3\mathit{ns}$ LHC extractions. The objective of this work is to probe that
the shape deforming effect due to spontaneous emission produces a backward
shift in time of $65.8\mathit{ns}$ theoretically calculated on the basis of
special relativity LASER equations. That is in agreement with the reported result.

\section{The Experiment}

First of all it is necessary to outline the experiment. The proton beam is
produced with the CERN Super Proton Synchrotron (SPS), these protons are
ejected with a kicker magnet, in two extractions each lasting $10.5\mathit{\mu
s}$ and separated by $50\mathit{ms}$ towards a graphite target where billions
of mesons are produced. These mesons are focused into a $1\mathit{km}$ vacuum
tunnel where the mesons decays into muons and neutrinos. Then neutrinos
continue traveling through the inside of the earth by $730\mathit{km}$ until
they arrive to Gran Sasso Laboratory $2.4\mathit{ms}$ later. Neutrinos are
detected by OPERA in two separated groups: the first with mean neutrino energy
of $13,9\mathit{GeV}$ and the second with $42,9\mathit{GeV}$, corresponding to
each one of the two proton extractions.

Proton extractions are similar to step functions with several peak or
oscillations superposed. These peaks corresponds to the proton synchrotron
radio frequencies of the SPS and the kicker magnet. The form of the time
distribution of proton function is accurately measured by a fast Beam Current
Transformer (BCT) at center of the graphite target. The key of the experiment
is that each maximum of the neutrinos detection must correspond to a maximum
of the proton intensity.

To obtain the time traveling of the neutrinos, each of $16000$ neutrino events
detected were tag in time, and correlated with its corresponding proton time
distribution function for each extraction with high accuracy. To do that, a
probability density function (PDF) is constructed, summing up all the proton
time distribution function, for which neutrino interactions were observed at
the detector. Then this function is shifted in time ($t\rightarrow t+TOF_{c}$)
by the estimated time of flight ($TOF_{c}=\frac{730.085\mathit{km}}{c}$) at
the speed of light $c$. The peaks of these PDF function must correspond in
time with the peaks in the neutrino detection if they fly exactly at the speed
of light. The measured neutrino time distribution, detected at OPERA must have
a delay time ($t\rightarrow t+TOF_{\nu}$), corresponding to the time of flight
of the neutrinos ($TOF_{\nu}=\frac{730.085\mathit{km}}{v}$) at its real
velocity $v$. As both time functions, the theoretical PDF function and the
detected neutrinos distribution had been shifted by $TOF_{c}$ and $TOF_{\nu}$
respectively; the maximum likehood analysis must gives the best fitting for a
time lapse \cite{OPERA}:%

\[
\delta t=TOF_{c}-TOF_{\nu}%
\]

If this time lapse is positive means that the neutrinos arrived faster than
expected for light, while if this quantity is negative means that the
neutrinos are slower than light.

To obtain the average delay in time, between the neutrino detection and the
PDF function, a numerical maximum likehood analysis is performed. The results
are shown separately for each of the the two extraction, that are enough
separated by $50\mathit{ms}$ to be uncorrelated. The results are summarized in
figure 1 taken from \cite{OPERA}:

\begin{figure}[h]
\includegraphics[width=14cm,angle=0]{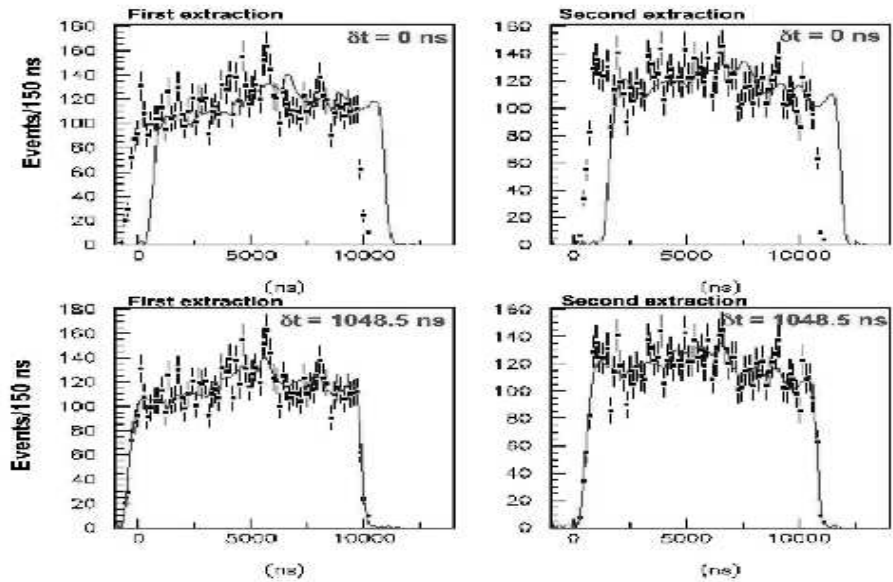}
\caption{Graphic of results of the PDF for protons and the detected neutrinos
detected by OPERA}%
\label{fig1}%
\end{figure}

Surprisingly, a positive lapse shift of $\delta t=1048.5\mathit{ns}$ was
obtained, indicating an early arrive than expected at the speed of light,
there are corrections due to the electronic time tag GPS, UTC, BCT, etc that
are summed up to $\ (985.8\pm7.4)\mathit{ns}$, when all the chain of systemic
errors were taken into account. So, there is still an unexplained forward
lapse shift of:
\[
(57.8\pm7.8\pm_{5.9}^{8.3})\mathit{ns}%
\]
that indicates that neutrinos are $0.25\times10^{-4}$ faster than known speed
of light in the vacuum. In this result had been included the statistical error
$\pm6.9\mathit{ns}$ obtained from the maximum likehood analysis, and checked
with various combination of MonteCarlo simulations.

Recently the OPERA collaboration has reported 20 neutrino events obtained for
very short proton extraction pulse, with a PDF gaussian mean width of only
$3\mathit{ms}$, in four bunches separated $524\mathit{ns}$ that is roughly one
thousand times shorter than former long events. This 20 neutrinos also appears
to precede light for
\[
62.1\pm3.7\mathit{ns}\text{ \ \ with a RMS: }16.4\mathit{ns}%
\]
These short extractions has a much lower particle density with only
$1.1\times10^{12}$ protons on target and mesons average lorentz factor as
large as $\gamma_{\pi}=190.$ This result of only 20 events has a very high
Root Mean Square of $RMS:=16.4\mathit{ns}$, so it will be significant only
when taken jointly with the previous $16000$ events result, but it has been
used to discard the shape deformation factor as a possible explanation to the
early neutrino arrival.

\section{Discussion.}

In a recent work \cite{Stimulated}, it has been proposed an explanation for
the neutrinos coming from the "long" $10500\mathit{ns}$ LHC\ extractions, that
apparently arrives to Gran Sasso $57.8\mathit{ns}$ earlier with respect to the
speed of light in vacuum. That apparent early arrival could be explained by a
shape deforming effect of the neutrino distribution function, with respect to
the proton distribution function (PDF) due to the stimulated absorption while
traveling by the decay tunnel that acts as a LASER tube, where other effects
as stimulated emission, absorption and spontaneous emission could take place.
In that work, it was obtained a shift factor for the maximum of the
distribution functions is obtained, due to the combined stimulated absorption
and emission effect:%

\begin{equation}
\Delta t=(N_{2}B_{21}-N_{1}C_{12})\sigma^{2}\label{deltaT}%
\end{equation}
where $N_{2}$ is the pion or meson density inside the decay tunnel, $N_{1}$is
the muon density, $B_{21}$is Einstein \cite{Einstein1917} stimulated emission
coefficient, $C_{12}$ is stimulated absorption coefficient and $\sigma$ the
gaussian root mean square of the initial neutrino pulse.

When stimulated absorption dominates over emission, there is negative or
backward shift in time for the maxima of the distribution functions while
traveling through the $L=1\mathit{km}$ decay tunnel. If only stimulated
absorption is considered (Beer-Lambert case, $B_{21}=0$) for the long pulse
extractions, with $N_{1}\approx10^{13}/vol$ (with a tunnel volume
$vol\approx3000\mathit{m}^{3}$) taking $\sigma\simeq3300\mathit{ns}$ (that is
also consistent with the tunnel length) with a stimulated absorption
coefficient:%
\[
C_{12}\simeq1.65\times10^{-6}\frac{\mathit{m}^{3}}{\mathit{\sec}}%
=5.49\times10^{13}\mathit{barn}\Rightarrow2.79\mathit{barn/muon}%
\]
will produce the $\approx60\mathit{ns}$ that had been reported by OPERA. But
for the LHC $\sigma=3\mathit{ns}$ short pulse extractions, with a density of
$1.1\times10^{12}/vol$, and the former $C_{12}$ coefficient will give an
insignificant time delay of%
\[
\Delta t=0.0000015\mathit{ns}%
\]

As equation(\ref{deltaT}) is strongly dependent on $\sigma^{2}$ that change
for a factor of $10^{6}$ (while density changes by an additional factor of 60)
stimulated absorption or emission could not explain the early arrival of short
and long CNGS beam simultaneously.

\section{Spontaneous emission in the context of special relativity}

It is not know in which part of the $1000\mathit{m}=3300\mathit{ns}$ tunnel
the decay of mesons into neutrino plus muon occurs, the statistical
correlation of thousands of neutrino peak distribution against the PDF
function will give the mean delay time with high statistical precision when it
is assumed that the spatial distribution of the decays in the tunnel is random
or gaussian, but it could happens that the starting point of neutrinos could
be shifted in time or driven by stimulated or spontaneous emission-absorption
processes, as happens for a pulse LASER traveling through an amplifier plasma
with an initial population \cite{Laser}.

The balance equation considering spontaneous emission ($A_{21}$ coefficient),
stimulated emission ($B_{21}$ coefficient) and stimulated absorption ($C_{12}$
coefficient) proposed in \cite{Stimulated}:%

\begin{align}
\frac{\partial n}{\partial t}+\frac{\partial n}{\partial x}.\nu &
=N_{2}\ A_{21}+N_{2}\ B_{21}\ n-N_{1}\ C_{12}\ n\label{laser1}\\
\frac{\partial n}{\partial t}+\frac{\partial n}{\partial x}.\nu &
=-\frac{\partial N_{2}}{\partial t}-\frac{\partial N_{2}}{\partial x}.v_{\pi
}\label{laser2}\\
\frac{\partial n}{\partial t}+\frac{\partial n}{\partial x}.\nu &
=\frac{\partial N_{1}}{\partial t}+\frac{\partial N_{1}}{\partial x}.v_{\mu}
\label{laser3}%
\end{align}
with neutrino density number $n$. The neutrino velocity is $\nu$, and the the
meson and muon velocities are $\nu_{\pi}$ and $\nu_{\mu}$ respectively. All
those densities are functions of the tunnel length $x$\ and the time $t$. For
short $\sigma=3\mathit{ns}$ LHC pulse extractions the $3300\mathit{ns}$\ decay
tunnel will be almost empty, so quadratic stimulated term could be a million
times less than spontaneous emission process, so the system of equations
reduces to:%

\begin{align}
\frac{\partial n}{\partial t}+\frac{\partial n}{\partial x}.\nu &
=N_{2}\ A_{21}\label{emission}\\
\frac{\partial N_{2}}{\partial t}+\frac{\partial N_{2}}{\partial x}.v_{\pi}
&  =-N_{2}\ A_{21} \label{absorption}%
\end{align}

The second equation is Lorentz invariant, but the first is no exactly. That is
due that for (\ref{absorption}) the 4-divergence is taken for a 4-volume
enclosed by the transverse area of the tunnel times a length equals $\nu_{\pi
}.\tau$ and multiplied by a time interval equals $\tau=(A_{21})^{-1}$ the mean
decay life of the meson. Under a Lorentz transformation, the time length
scales a Lorentz factor while the spatial length scale by the inverse of the
Lorentz factor then the volume in which is taken (\ref{absorption}) is
invariant. But for equation (\ref{emission}), the volume length equals
$\nu.\tau$, where the speed of neutrinos is involved instead of that of the
mesons, also the time interval to be taken is $\tau=(A_{21})^{-1}$, where the
mean decay time life is measured in the rest or Lab system and not as seen
from the meson. To correct this difference in the volumes in which particles
are destroyed or created a factor of $v/v_{\pi}$ must be included but at the
rest or Lab system this factor is almost one: $N_{2}\ A_{21}.v/v_{\pi}\simeq
N_{2}\ A$, but equation (\ref{emission}) is valid only in the Lab system
reference frame.

So in order to get the correct solution to the process we must go to the meson
reference frame where $v_{\pi}=0$ and
\[
\nu\rightarrow c_{v}^{\prime}=\frac{\nu-\nu_{\pi}}{1-\frac{\nu\nu_{\pi}}%
{c^{2}}}%
\]
then the equation system is:%

\begin{align}
\frac{\partial n^{\prime}}{\partial t^{\prime}}+\frac{\partial n^{\prime}%
}{\partial x^{\prime}}c_{v}^{\prime}  &  =\frac{N_{2}^{\prime}}{\tau_{o}%
}\label{una}\\
\frac{\partial N_{2}^{\prime}}{\partial t^{\prime}}  &  =-\frac{N_{2}^{\prime
}}{\tau_{o}} \label{dos}%
\end{align}
where $\tau_{o}=26\mathit{ns}$ is the mean life of the Pion in the meson
reference frame.

Equation (\ref{dos}) could be easily integrated%
\begin{align}
N_{2}^{\prime}  &  =N_{20}^{\prime}(x^{\prime})\exp(-\frac{t^{\prime}}%
{\tau_{o}})\label{N2prima}\\
N_{2}^{\prime}  &  =N_{20}^{\prime}\exp(-\frac{(x^{\prime}/\nu_{\pi})^{2}%
}{2\sigma_{o}^{2}})\exp(-\frac{t^{\prime}}{\tau_{o}}) \label{N2exp}%
\end{align}
where a gaussian shape as in \cite{Stimulated} was used and $\sigma_{o}$ is in
the meson reference frame. Equation (\ref{una}) could also be integrated in
term of the error function, but is easier to boost back to the Lab reference frame.%

\begin{align}
\frac{x^{\prime}}{\nu_{\pi}} &  \rightarrow(\frac{x}{\nu_{\pi}}-t)\gamma_{\pi
}=-\xi_{\pi}\gamma_{\pi}\approx-\xi\gamma_{\pi}\text{ }\label{xprima}\\
t^{\prime} &  \rightarrow(t-\nu_{\pi}x)\gamma_{\pi}=\frac{1}{2}[\xi_{\pi
}\gamma_{\pi}(1+\nu_{\pi}^{2})+\frac{\eta_{\pi}}{\gamma_{\pi}}]\approx
\xi\gamma_{\pi}+\frac{\eta}{2\gamma_{\pi}}\label{tprima}\\
\eta_{\pi} &  =t+\frac{x}{\nu_{\pi}},\;\eta=t+x,\\
\xi_{\pi} &  =t-\frac{x}{\nu_{\pi}},\;\xi=t-x
\end{align}
where the approximation $v_{\pi}^{2}\approx1$ has been used at right hand
equations. Then the solution to (\ref{una}) and (\ref{dos}) could be written
in the Lab Reference frame as a function of light coordinates $(\eta,\xi)$
instead of the more exact characteristic coordinates $(\eta_{\pi},\xi_{\pi})$
\begin{align}
N_{2} &  =N_{20}\exp(-\frac{\xi^{2}}{2\sigma^{2}})\exp(-\frac{\eta}{2\tau
})\exp(-\frac{\xi\gamma_{\pi}}{\tau_{o}})\label{N2lab}\\
n &  =N_{20}\exp(-\frac{\xi^{2}}{2\sigma^{2}})\exp(-\frac{\xi\gamma_{\pi}%
}{\tau_{o}})[1-\exp(-\frac{\eta}{2\tau})]\label{nlab}%
\end{align}
where $\sigma=\sigma_{o}/\gamma_{\pi}$ and $\tau=\gamma_{\pi}\tau_{o}$ are in
reference to the Lab Frame.

Proceeding as in \cite{Stimulated} looking for the maximum:%

\begin{align}
\frac{\partial n}{\partial t} &  =\frac{\partial n}{\partial\xi}%
+\frac{\partial n}{\partial\eta}=0\\
0 &  =\left[  (-\frac{\xi}{\sigma^{2}}-\frac{\gamma_{\pi}}{\tau_{o}}%
)[1-\exp(-\frac{\eta}{2\tau})]+\frac{1}{2\tau}\exp(-\frac{\eta}{2\tau
})\right]  _{(t_{2},L)}\label{maximo}%
\end{align}
then, valuating at the tunnel's end $[\xi]_{(t_{2},L)}=\Delta t=t_{2}-t_{1}$
and $[\eta]_{(t_{2},L)}=\Delta t+2L/c$ where $L=1\mathit{km}$ \ and
$c=3\times10^{5}\mathit{km/s}$. Equation (\ref{maximo}) could be rewritten as:%
\begin{equation}
\frac{1}{2\tau}=\left(  \frac{\Delta t}{\sigma^{2}}+\frac{\gamma_{\pi}}%
{\tau_{o}}\right)  \exp[\frac{\Delta t}{2\tau}](1+\exp[\frac{L}{\tau
c}])\label{trascendent}%
\end{equation}
taking into account that $\Delta t\approx-60\mathit{ns}<<2\tau=9880\mathit{ns}%
$, \ then $\exp[\frac{\Delta t}{2\tau}]\approx1$\ in order to solve the
transcendent equation (\ref{trascendent}) using that $(1+\exp[\frac{L}{\tau
c}])\approx2.95$ equation (\ref{trascendent}) could be solved by%

\begin{equation}
\Delta t=-\frac{\gamma_{\pi}\sigma^{2}}{\tau_{o}}+\frac{\sigma^{2}}%
{5.9\gamma_{\pi}\tau_{o}}=-65.8\mathit{ns}+0.0003\mathit{ns}\label{deltaShort}%
\end{equation}

As the forward time shift, for the 20 short pulse neutrinos obtained for
OPERA\ is $-62.1\pm3.7\mathit{ns,}$ the theoretical obtained $\Delta
t=-65.8\mathit{ns}$ completely explain it. This striking result make us to
ask, if also the $57.8\mathit{ns}$ early arrival for the "long"
$10500\mathit{ns}$ \ pulse extractions event could be explained by the
$200\mathit{Mhz}$ LHC harmonic, that is superposed to the long pulse with a
period of $T=5\mathit{ns}$, as could be seen in the following picture taken
from \cite{OPERA}.

\begin{figure}[h]
\includegraphics[width=12cm,angle=0]{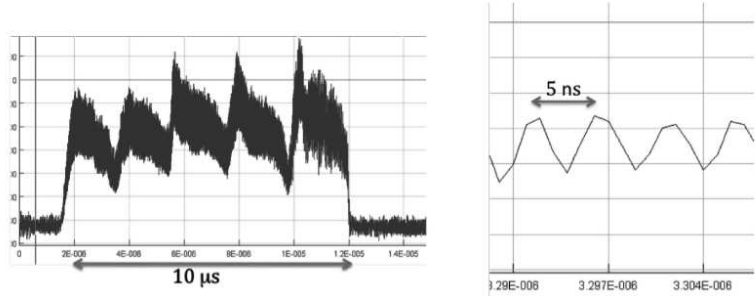}
\caption{Graphic of results of the PDF for the protons from LHC showing  the
200Mz harmonic.}%
\label{fig2}%
\end{figure}

If we approximate this harmonic in the interval $[-\frac{T}{2},\frac{T}{2}]$
by a gaussian function
\[
\frac{1}{2}(1+\cos\frac{2\pi\hat{t}}{T})\approx\exp(-\frac{\hat{t}^{2}}%
{2\hat{\sigma}^{2}})
\]
for $\hat{t}=$ $\hat{\sigma}$ we get: $\hat{\sigma}=1.08\mathit{ns}$ that is
enough to explain a early arrival of only
\[
\Delta\hat{t}=-\frac{\gamma_{\pi}\hat{\sigma}^{2}}{\tau_{o}}=-7.3\mathit{ns}%
\]

So apparently, spontaneous emission only, could not explain the early arrival
of long CNGS pulses, but as was done in \cite{Stimulated} the remaining
$50\mathit{ns}$ could be fully explained if stimulated absorption (and
possible stimulated emission) is taken into account.

\section{Conclusions}

In conclusion the deformation of the $\sigma=3\mathit{ns}$ short pulses of
neutrinos, calculated with relativistic corrections, produce an
"\textit{apparent"} earlier arrival of the neutrinos, with respect to the
speed of light by%

\[
\Delta t=-\frac{\gamma_{\pi}\sigma^{2}}{\tau_{o}}=-65.8\mathit{ns}%
\]
explaining the reported earlier arrival of$\ -62.1\pm3.7\mathit{ns}%
$\ \ completely. So the short pulse neutrinos are not superluminal, there is a
relativistic time shift that produce a deformation of the wave distribution
function that shift the maximum, so it\ will arrive earlier, not real
particles. Here is a clear prediction: the time shift for short pulses will be
linear in the Lorentz factor and quadratic in the mean gaussian width $\sigma$
that could easily tested at OPERA with the actual infrastructure.

As a second conclusion, the spontaneous emission process alone could not
explain the $-57.8\mathit{ns}$. But it could be fully explained if stimulated
absorption-emission processes were taken into account. The calculation
performed in \cite{Stimulated} is a very simple and restricted approximation,
exact solutions of (\ref{laser1})(\ref{laser2}) and (\ref{laser3}) that takes
into account relativistic correct volumes and real pion, muon and neutrino
velocities are required.

There is a way to probe the existence of these shape deformation processes: to
compare the detected neutrino time distribution with the muon probability
distribution function, instead of the meson probability distribution function,
because as the muon and the neutrino are created at the same event: they must
have the same forward time shift so there, must be not difference in time at
which intensities maximum are achieved. In \cite{OPERA} the muon PDF was not
considered, because the simulations show that will give null corrections. That
is not true as was shown in this paper. Muon PDF may be difficult to analyze
because the muon is a very interactive charged particle, but comparison of
muons PDF with detected neutrinos must be performed in order to establishes if
neutrinos travel faster than light or stimulated absorption or emission of
neutrinos exist.




\begin{thebibliography}{9}                                                                                                %


\bibitem {OPERA}The OPERA Collaboration, ``Measurement of the neutrino
velocity with the opera detector in the CNGS beam'\ (2011), arXiv:1109.4897v2.

\bibitem {1987A}Arnett, W.D.; et al. (1989). "Supernova 1987A". Annual Review
of Astronomy and Astrophysics 27:629--700; K. S. Hirata, et al, ``Observation
in the Kamiokande-II detector of the neutrino burst from supernova
SN1987A'\ Physical Review D 38, 448--458 (1988).

\bibitem {Laser}C. Rulliere (ed.). "Femtosecond Laser Pulses: Principles and
Experiments", [2nd ed.]. Springer (2005).

\bibitem {Einstein1917}A. Einstein, "Strahlungs-Emission und Absorption nach
der Quantentheorie". Verh. der Physikal. Ges.18 (1916)318; A. Einstein, "On
the quantum theory of radiation". Source of Quantum Mechanics. Classical of
Science Volume V. Dover Publication (1917)

\bibitem {Glashow}A. Cohen, S. Glashow,"New constraint on neutrino
velocities". arXiv:1109.6562. Xiao-Jun Bi, Peng-Fei Yin, Zhao-Huan Yu, Qiang
Yuan,"Constraints and tests of the OPERA superluminal neutrinos". arXiv:1109.6667.

\bibitem {ICARUS}ICARUS collaboration: "A search for analogue to Cherenkov
radiation by high energy neutrinos at superluminal speeds in Icarus". arXiv 1110.3763v2[hep-ex].

\bibitem {Stimulated}Rafael Torrealba, "Using an Einstein's idea to explain
OPERA faster than light neutrinos". arXiv:1110-0243v3[hep-ph].
\end{thebibliography}
\end{document}